\documentclass{JFM-FLM_Au}

\usepackage{graphicx}
\usepackage{caption}
\usepackage{subcaption}

\lefttitle{N.A. DeFilippis, O. B\"uhler, K.S. Smith}
\righttitle{Journal of Fluid Mechanics}

\title{Frequency spreading of internal wave energy by balanced flows in two dimensions}

\author{Nicholas DeFilippis\aff{1}, Oliver B\"uhler\aff{1} \and K. Shafer Smith\aff{1}}

\affiliation{\aff{1}{Courant Institute of Mathematical Sciences, New York University, NY 10012, USA}}

\corresau{Nicholas DeFilippis, \email{nad9961@nyu.edu}}

\begin{document}
\maketitle

\begin{abstract}

Interactions between inertia-gravity waves and balanced flows lead to a spectral diffusion of wave action. Prior work has established that this diffusion is weak across constant frequency surfaces in three-dimensional settings, but can be significant in two dimensions with a non-stationary balanced flow. We investigate the two-dimensional setting through numerical simulations that simultaneously evolve a turbulent quasigeostrophic balanced flow and advect rotating shallow water wave packets. In contrast to earlier predictions based on the synthetic flows used by Dong \textit{et al.} (\textit{J. Fluid Mech.}, 2020, vol. 905, R3), we find that frequency spreading from wave mean-flow interactions is weaker for realistic turbulent flows. We derive a timescale for frequency diffusion and show that frequency spreading with a realistic background flow is an order of magnitude smaller than with the synthetic flow. We narrow the discrepancy between the two- and three-dimensional induced diffusion theories, which suggests other mechanisms are responsible for the broadband frequency spectra seen in the atmosphere and ocean.

\end{abstract}

\begin{keywords}
Authors should not enter keywords on the manuscript, as these must be chosen by the author during the online submission process and will then be added during the typesetting process (see \href{https://www.cambridge.org/core/journals/journal-of-fluid-mechanics/information/list-of-keywords}{Keyword PDF} for the full list).  Other classifications will be added at the same time.
\end{keywords}


\section{Introduction}
\label{sec:introduction}


The interactions between inertia-gravity waves and balanced flows in rotating stratified fluids are weak, but their ubiquity can lead to significant effects over long times. \cite{kafiabad_diffusion_2019} showed, for example, that in a weak-flow limit with WKB scale separation, a frozen balanced flow catalyzes a cascade of internal wave energy across scales through the ``induced diffusion'' of wave action \cite{ctx40894637160007876}.  Investigated in a 3D Boussinesq setting, their results offer a compelling explanation for observations of energy spectra in the atmosphere \citep{TheoreticalInterpretationofAtmosphericWavenumberSpectraofWindandTemperatureObservedbyCommercialAircraftDuringGASP} and ocean \citep{InterpretingEnergyandTracerSpectraofUpperOceanTurbulenceintheSubmesoscaleRange1200km}.

Notably, in this work diffusion spreads action freely along the cone of constant intrinsic frequency, but weakly across it through conservation of absolute frequency. This suggests that action diffusion by mean flows cannot explain the ubiquitous broadband oceanic internal wave frequency spectrum \citep{garrett_space-time_1972}. Interested in the possibility, \cite[][hereafter DBS]{Dong_Bühler_Smith_2020}  revisited this problem in the two-dimensional rotating shallow water setting, showing that, theoretically at least, relaxing the frozen mean flow assumption allows for action diffusion across constant-frequency surfaces, as a time-dependent mean flow allows for absolute frequency to vary in time as well. In two dimensions, for small Froude number mean flows, time dependence is required for a significant change in intrinsic frequency, as it is closely tied to absolute frequency. They demonstrated the effect numerically using integrations of the ray tracing equations for rotating shallow waves --- where a cascade in wavenumber space cannot occur without also spreading wave action in intrinsic frequency space --- in a synthetic balanced flow. Time-dependence in the flow was generated by modeling each flow wavenumber as an Ornstein–Uhlenbeck process, with variable decorrelation times. Integrating rays in flows with a range of decorrelation times, it was found that the radial action diffusivity was non-zero, with a peak that scales like Froude number to the third power. 

\cite{dong_geostrophic_2023} returned to the 3D Boussinesq case to explore how wave capture \citep{Buhler_McIntyre_2005} might spread wave action across cones of constant frequency. They found significant spreading of intrinsic frequencies for both time-dependent and time-independent background flows based on the mechanism of wave capture, which takes over after the diffusion regime breaks down in three-dimensional flow.

In light of these results, \cite{Cox_Kafiabad_Vanneste_2023} considered in great detail whether time-dependence of the mean flow could lead to spreading of wave action across frequencies in the 3D Boussinesq system in the induced diffusion regime, concluding theoretically that it is still weak. Their conclusion is buttressed by convincing high-resolution simulations of the Boussinesq system. They explain that the difference in results between the two-dimensional shallow water waves in DBS and the 3D Boussinesq waves is due to the compactness and non-compactness, respectively, of constant frequency surfaces.

Further understanding the two-dimensional results of DBS suggests exploring wave frequency spreading in a direct numerical simulation of the RSW equations. One issue with the RSW system, however, is its formation of unphysical shocks. \cite{thomas_turbulent_2024} shows that such shocks drive a forward cascade of wave energy in wavenumber space (\cite{augier2019shallow} show a similar result, albeit for the nonrotating RSW system), suggesting that their presence may obscure the possible role of balanced eddies in catalyzing a wave energy cascade. In order to correct for this potential issue, we perform simulations using both the RSW, and a modified version that replaces the pressure term with a nonlinear formulation that prevents wave-steepening, but leaves the potential vorticity and linear dynamics unaltered \citep{AShallowWaterModelthatPreventsNonlinearSteepeningofGravityWaves}. Consistent with the findings of \cite{Cox_Kafiabad_Vanneste_2023}, we find only minimal wave energy spreading in the modified equations.

This leads to the question, to what extend does frequency diffusion depend on the specific synthetic background flow chosen in \cite{Dong_Bühler_Smith_2020}? Here, in addition to the DNS experiments, we reconsider the DBS approach by employing quasigeostrophic simulations to generate balanced flows that simultaneously advect RSW wave packets. We develop multiple methods by which to compute the radial wavenumber diffusion of wave action.  Ultimately, we find that realistic balanced eddy fields result in an action diffusivity that scales more steeply with Froude number than that suggested by the synthetic flow fields used in DBS.

Section \ref{sec:shallow-water} captures the complete two-dimensional picture with a direct numerical simulation (DNS) of the rotating shallow water equations. These results reveal an apparent contradiction between the DBS results, which see strong frequency spreading for time-dependent flow, and the inherently time-dependent DNS that only sees weak frequency spreading. Section \ref{sec:raytracing} discusses the ray tracing method and numerical simulations, which only permit wave-mean flow interactions. These simulations reveal a steep dependence between the Froude number of the mean flow and the degree of frequency spreading. Section \ref{sec:analysis} explicitly quantifies the dependence of frequency spreading on the Froude number of the mean flow in a few ways, including by analogy to geometric Brownian motion. This result narrows the gap between the lack of spreading seen in three-dimensional flows and the strong spreading seen in two dimensions.

\section{Direct simulation of wave-frequency spreading in the rotating shallow water model}
\label{sec:shallow-water}


We begin by investigating wave energy spreading using direct numerical simulations of two flavours of the rotating shallow water equations. A direct numerical simulation (DNS) captures all nonlinear dynamics, including wave-mean flow interactions, providing a baseline to which reduced models can be compared. For small-amplitude flow, the rotating shallow water equations permit a meaningful linear decomposition into geostrophic and wave components. However, the standard rotating shallow water equations produce shocks with broad spectral footprints, obscuring the true energy spectrum. To avoid these effects, we also integrate a modified rotating shallow water model that prevents shocks \citep{AShallowWaterModelthatPreventsNonlinearSteepeningofGravityWaves}. Simulations of the standard and modified equations are compared to isolate the role of shocks in spreading wave energy from large to small scales.

\subsection{The rotating shallow water equations with and without shocks}
\label{sec:dns}

We consider the rotating shallow water equations for a fluid of constant mean depth $H$ with dissipation,
\begin{subequations}\label{eq:rsw}
    \begin{gather}
    \p_t \boldsymbol{u} + (\boldsymbol{u}\bcdot\bnabla)\boldsymbol{u} + f\hat{\boldsymbol{z}}\times\boldsymbol{u} = -c^2 \bnabla\eta + \mathcal{D} \boldsymbol{u},\label{eq:rsw-momentum}\\
    \p_t\eta + \bnabla\bcdot\left[(1 + \eta)\boldsymbol{u}\right] = \mathcal{D}\eta,
    \end{gather}
\end{subequations}
where $\boldsymbol{u}=(u,v)$ is the horizontal velocity vector, $\eta$ is the non-dimensional deviation from the mean height of the fluid, $f$ is the Coriolis parameter, $\hat{\boldsymbol{z}}$ is the vertical unit vector, $c=\sqrt{gH}$ is the non-dispersive wave speed, and $\mathcal{D}=-\nu\nabla^8$ with $\nu>0$ is a hyperviscous dissipation operator.

The shallow water equations are well known to produce shocks, even with nonzero rotation $f\neq0$ \citep{thomas_turbulent_2024}. These shocks have a distinct spectral signature that typically occludes the degree of wave spreading. To quantify this effect, we implemented both \eqref{eq:rsw} and a modified version that had been developed in \cite{AShallowWaterModelthatPreventsNonlinearSteepeningofGravityWaves} to prevent shock formation. The modified equations replace the pressure term in \eqref{eq:rsw-momentum} via
\begin{equation}\label{eq:msw}
\bnabla\eta \quad\rightarrow\quad \bnabla\left(-\frac{1}{2(1 + \eta)^2}\right) = \bnabla\left(\eta -\frac{3}{2} \eta^2 + \ldots\right).
\end{equation}
This modification removes the shock-forming tendency but  does not affect potential vorticity~(PV) conservation or the linear structure of the equations.

The linear PV is $q=\zeta-f\eta$, where $\zeta=v_x-u_y$. Standard and modified equations have the same linear PV and quasigeostrophic dynamics. The linearised form of equations \eqref{eq:rsw} can be decomposed into a divergence-free geostrophic mode and two wave modes with zero linear PV \citep{SalmonRick1998Logf}, where the wave modes satisfy the rotating shallow water dispersion relationship
\begin{equation}\label{eq:shallow-water-dispersion}
    \omega(\boldsymbol{k}) = +\sqrt{f^2 + c^2\vert\boldsymbol{k}\vert^2},
\end{equation}
where $\omega(\boldsymbol{k})\geq0$ is the intrinsic frequency of the wave. We do not consider the effects of topography or mean height variation. These effects are treated in \cite{Cox_Kafiabad_Vanneste_2025}. This decomposition is applied for small-amplitude flow to approximate the geostrophic and wave components of the non-linear equations. 

\subsection{Direct numerical simulations}

We perform direct numerical simulations of both the rotating shallow water equations \eqref{eq:rsw} and their modified variant \eqref{eq:msw}, using a pseudo-spectral scheme implemented in the FourierFlows.jl Julia framework \citep{navid_c_constantinou_2025_17281674} on a $2\upi\times 2\upi$ periodic domain with $1024\times 1024$ grid points. This corresponds to an effective maximum wavenumber of 384 using a $1/3$-dealiasing scheme. The numerical integration uses an Adams-Bashforth third-order integration scheme with an integration factor method, which allows for an exact solution to the linear terms. The time step $\delta t$ is chosen to satisfy the Courant-Friedrichs-Lewy (CFL) condition. The dissipation coefficient, $\nu$, is chosen such that enstrophy dissipation is independent of the domain resolution and the amplitude of the flow. With an 8th-order dissipation scheme, and assuming a constant CFL condition, this requires $\nu \sim \delta t^{-1}K_{\textrm{max}}^8$, where $K_{\textrm{max}}$ is the maximum resolved wavenumber at a given domain size and resolution.

The Coriolis parameter and nondispersive wavespeed are set to $f=3$ and $c=1$, resulting in a deformation wavenumber $K_D = f/c = 3$.  The initial flow condition is separated into geostrophic and wave components, using the wave-geostrophic decomposition and random phases. To simulate near-inertial oscillations interacting with mesoscale eddies, the wave energy is initialised in the near-inertial waveband $(0, \frac{5}{3}K_D)$, as if disturbed by a large-scale ocean storm. The geostrophic energy is initially concentrated in $(\frac{10}{3}K_D, \frac{13}{3}K_D)$ to represent mesoscale eddies. We scale the wave components such that the initial root-mean-square (rms) wave velocity is $\langle \boldsymbol{u}_\text{wave}\rangle = 0.04$, and the geostrophic components so that the initial rms balanced velocity is $\langle \boldsymbol{u}_\text{balance}\rangle = 0.07$.  The simulation is spun up to $ft = 400$ and then analysed up to $ft = 433$. Using the averaged balanced rms velocity over this analysis period, the Froude number is $Fr = 0.044$, where
\begin{equation}\label{eq:Fr}
\textup{Fr}=\langle\boldsymbol{u}_\text{balance}\rangle/c.
\end{equation}

\begin{figure}[t]
    \centering
        \begin{subfigure}[t]{0.48\linewidth}
            \centering
            Standard shallow water
            \begin{subfigure}[t]{0.48\linewidth}
                \centering
                \includegraphics[width=\textwidth]{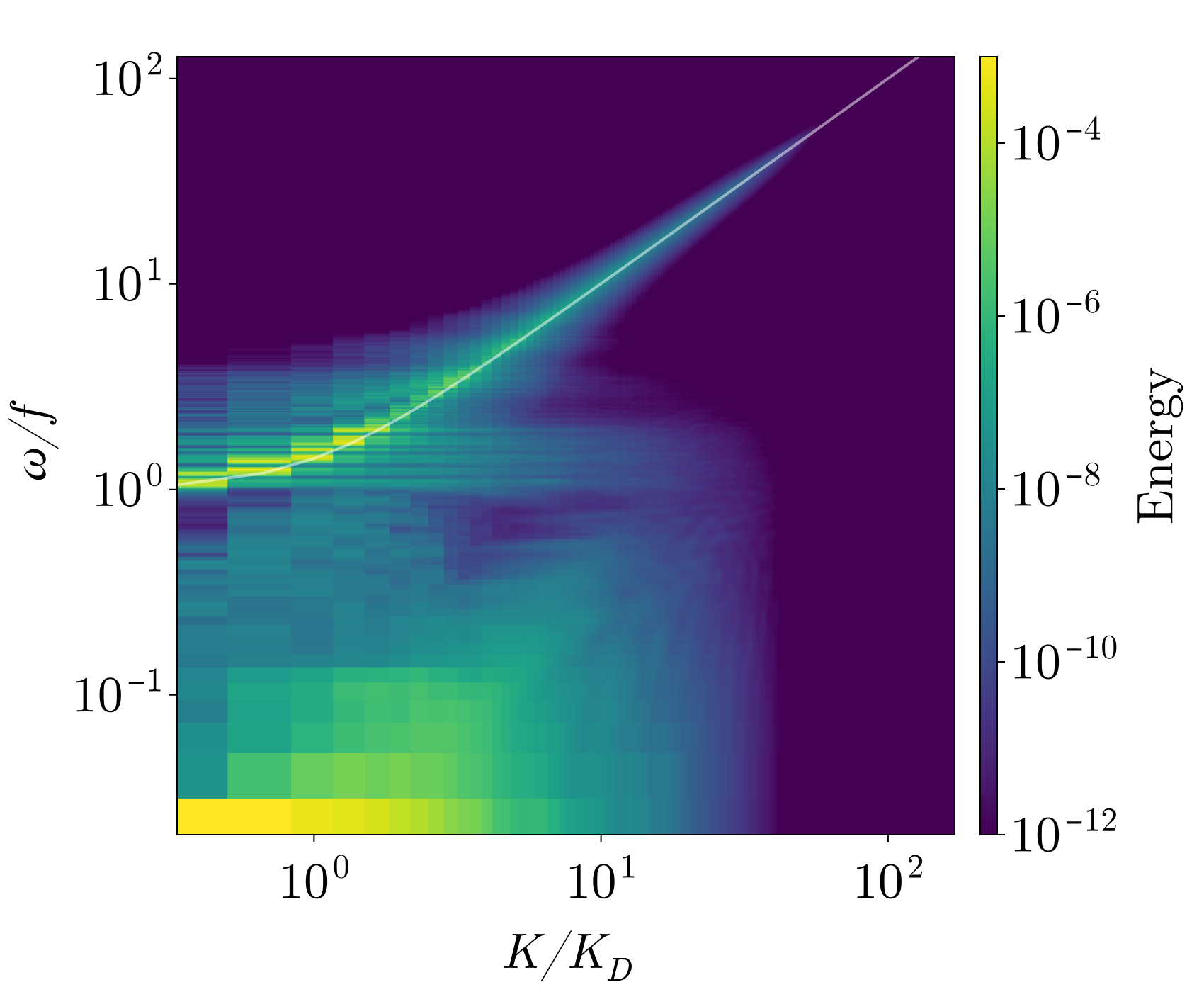}
                \caption{}
                \label{fig:rsw-komega}
            \end{subfigure}
            \begin{subfigure}[t]{0.48\linewidth}
                \centering
                \includegraphics[width=\textwidth]{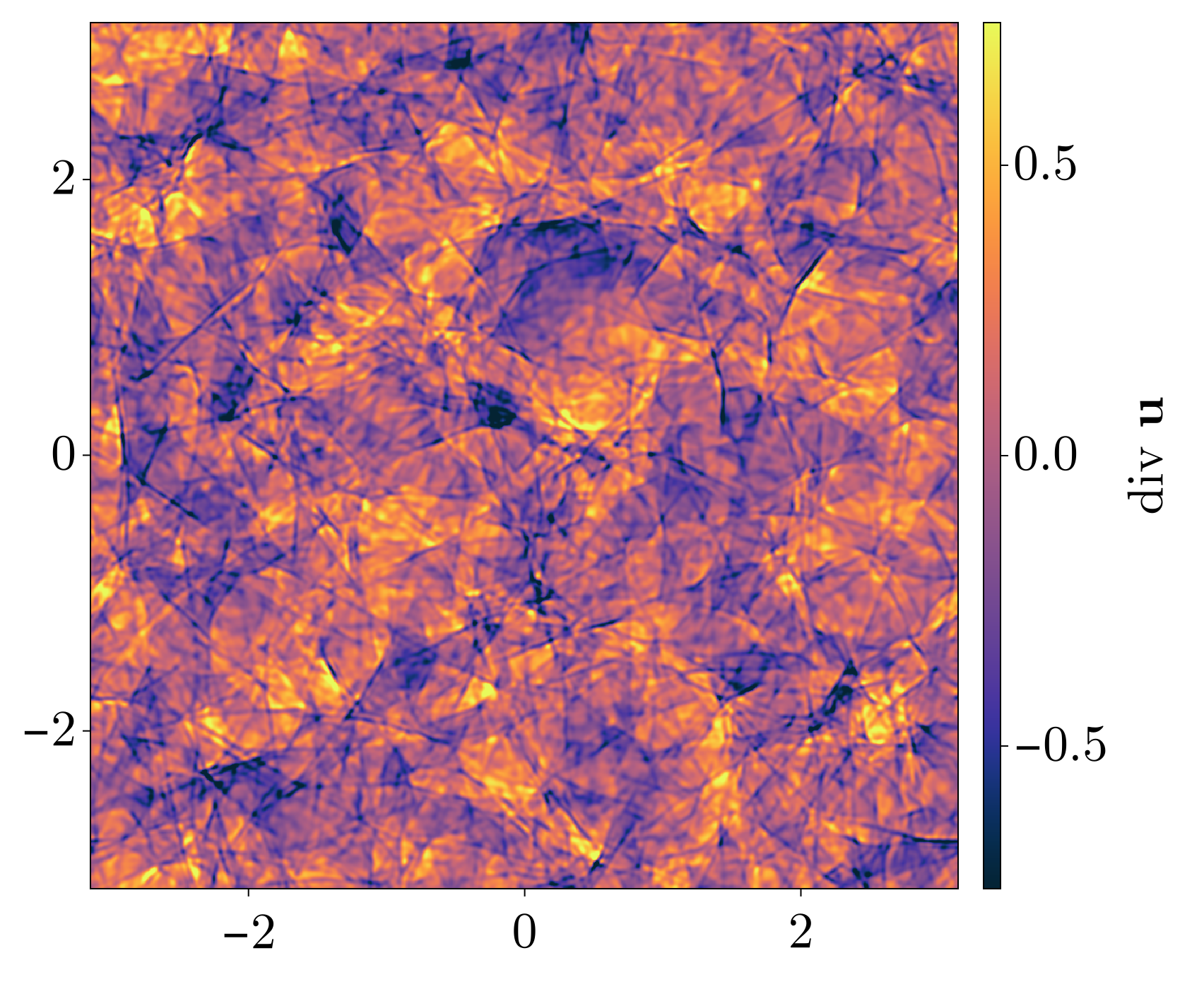}
                \caption{}
                \label{fig:rsw-div}
            \end{subfigure}
        \end{subfigure}
        \begin{subfigure}[t]{0.48\linewidth}
            \centering
            Modified shallow water
            \begin{subfigure}[t]{0.48\linewidth}
                \centering
                \includegraphics[width=\textwidth]{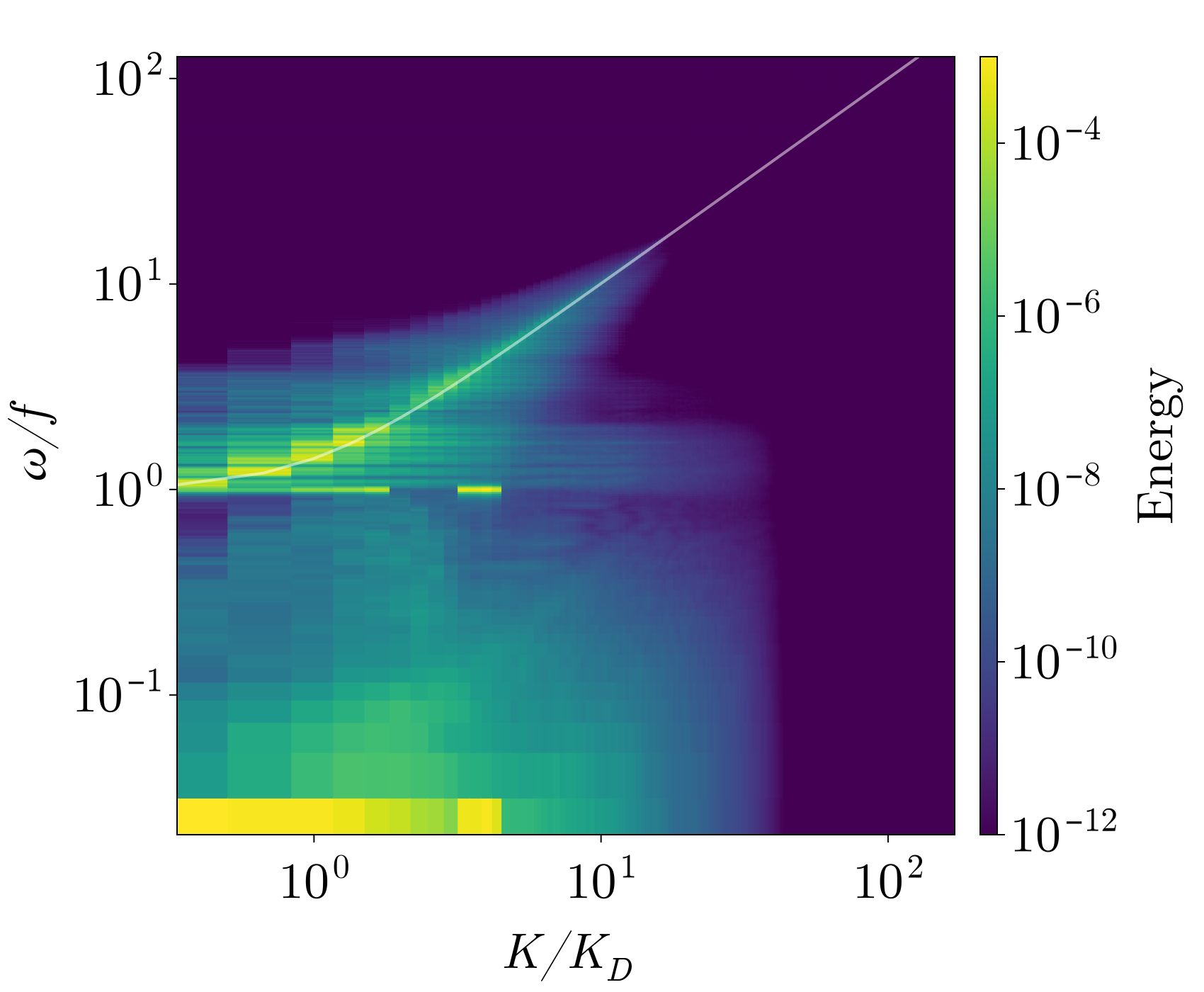}
                \caption{}
                \label{fig:msw-komega}
            \end{subfigure}
            \begin{subfigure}[t]{0.48\linewidth}
                \centering
                \includegraphics[width=\textwidth]{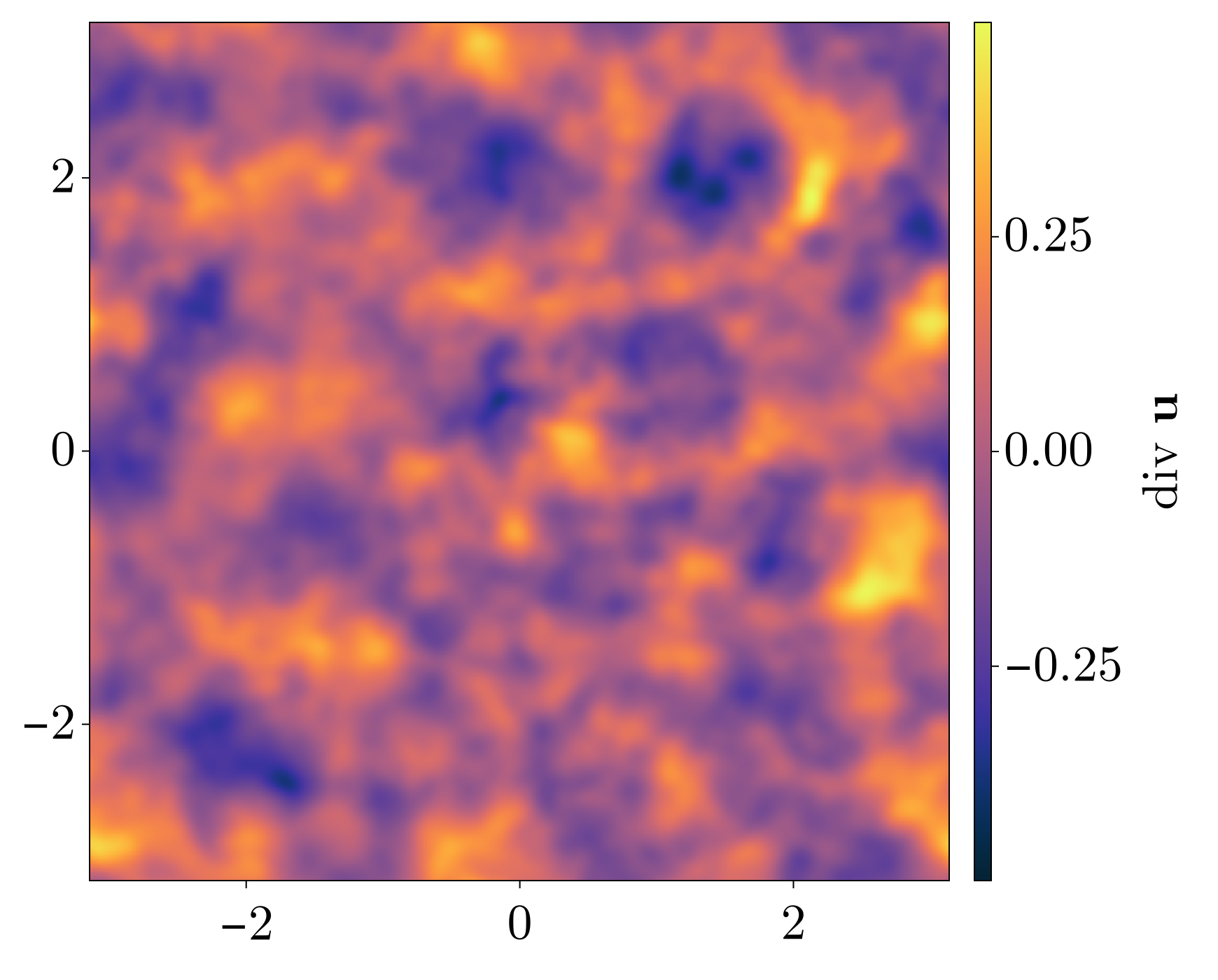}
                \caption{}
                \label{fig:msw-div}
            \end{subfigure}
        \end{subfigure}
        \begin{subfigure}[t]{\linewidth}
            \centering
            \includegraphics[width=0.5\textwidth]{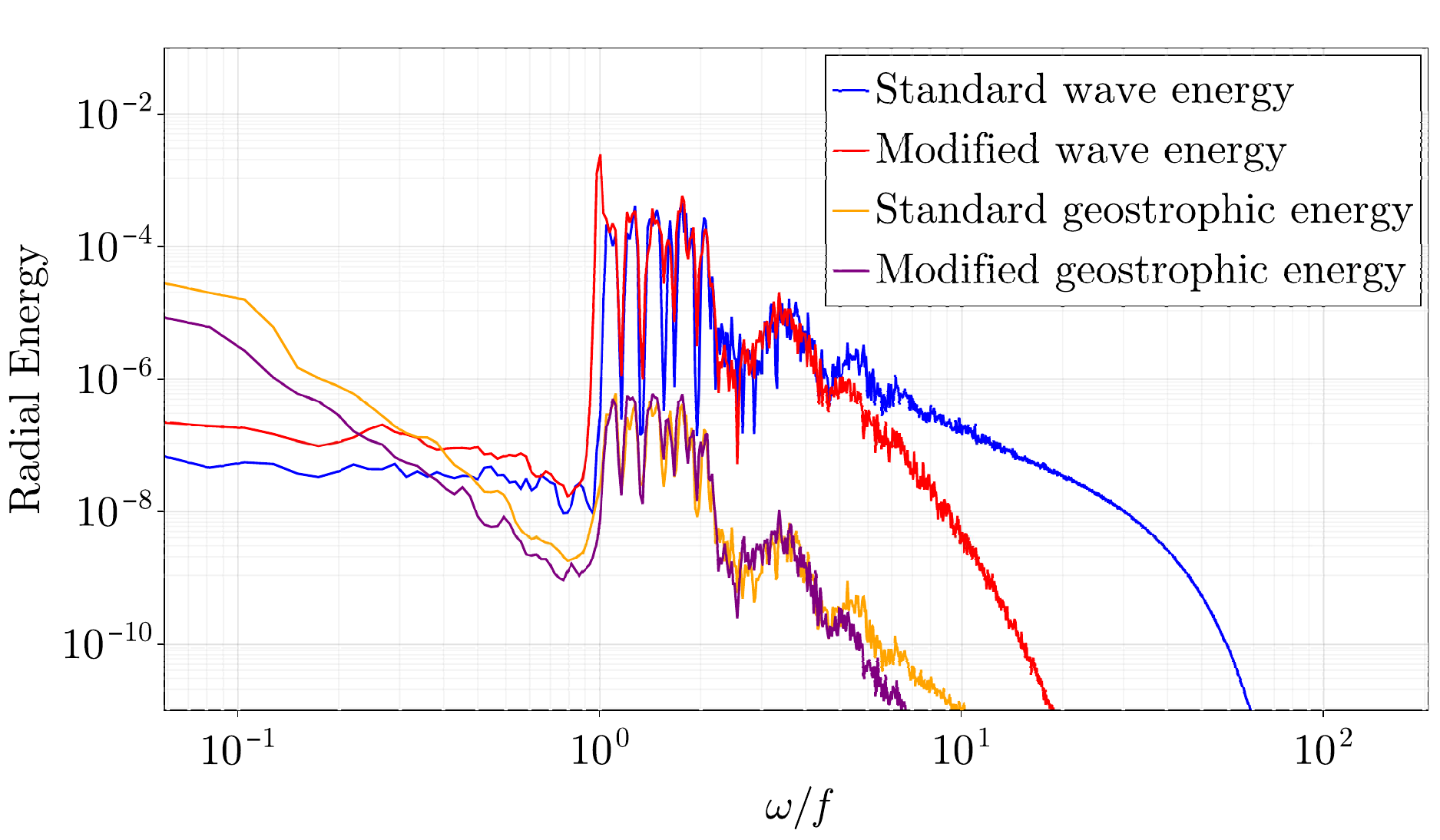}
            \caption{}
            \label{fig:sw-spec}
        \end{subfigure}
    \caption{Standard \eqref{eq:rsw} vs modified  \eqref{eq:msw} shallow water simulations. (\protect\ref{fig:rsw-komega}), (\protect\ref{fig:msw-komega}) Wavenumber-frequency space spectra for standard and modified shallow water, respectively, with the shallow water dispersion relation overlaid with a white line. The standard equations exhibit more spreading along this line. (\protect{\ref{fig:rsw-div}}), (\protect{\ref{fig:msw-div}}) Divergence field of standard and modified shallow water. Even with a rotational component, the standard shallow water equations still display scar-like shock features. On the other hand, these features are absent in the divergence of the modified equations. It is these scars that lead to an artificially shallower wave spectrum for the standard equations compared to the modified equations. (\protect{\subref{fig:sw-spec}}) Marginal distribution in frequency of the energy spectrum for the wave and geostrophic components of each equation. While the geostrophic energies for the standard and modified (orange and purple lines, respectively) are similar, the wave energy of the modified equation (red line) is much steeper and extends to smaller frequencies than the standard equations (blue line).}
    \label{fig:rsw-msw-comparison}
\end{figure}

The standard and the modified shallow water simulations are compared in Figure~\ref{fig:rsw-msw-comparison}.  Frequency-wavenumber spectra (panels \ref{fig:rsw-komega} and \ref{fig:msw-komega}) both show energy distributed along the shallow-water dispersion curve, but more so in the standard shallow water model. Incipient shocks can be seen in the standard shallow water simulation, where they form scar lines of strong negative divergence (panel \ref{fig:rsw-div}). By contrast, the modified equations do not exhibit these shocks (panel \ref{fig:msw-div}). While the geostrophic energy between the two simulations is comparable, the wave energy of the modified equations has a much steeper frequency slope and quickly drops off from its initial distribution (panel \ref{fig:sw-spec}). This suggests that the influence of wave-mean flow interactions are not as strong as previously thought in creating a broadband frequency spectrum. 

\section{Shallow water waves with a quasigeostrophic mean flow}\label{sec:raytracing}

The lack of significant frequency diffusion in direct numerical simulations of the modified rotating shallow water equations seems inconsistent with the vigorous spreading found in the ray tracing simulations of DBS. Among their differences, the full shallow water simulations capture all nonlinear interactions, while the ray tracing approach omits wave-wave interaction, and moreover assumes a spatially and temporally slowly-varying mean flow. Additionally, DBS used a synthetically-generated time-varying mean flow intended to mimic a balanced geostrophically-turbulent velocity field. Here we revisit this approach, upgrading the synthetic mean flow with a velocity field computed (simultaneously with the ray tracing simulation) from simulations of stationary two-layer quasigeostrophic turbulence.

\subsection{Ray tracing simulations}

%

The ray tracing approach represents waves as a set of distinct packets, each defined by its position vector $\boldsymbol{x}$ and wavenumber vector $\boldsymbol{k}$. Their evolution is determined by 
\begin{subequations}\label{eq:raytracing}
    \begin{align}
        \dot{\boldsymbol{x}} &=\boldsymbol{U}(\boldsymbol{x},t)+\bnabla_{\boldsymbol{k}}\omega(\boldsymbol{k}),\\
        \dot{\boldsymbol{k}} &= -\left[\bnabla_{\boldsymbol{x}}\boldsymbol{U}(\boldsymbol{x},t)\right]\bcdot\boldsymbol{k},
     \end{align}
\end{subequations}
where $\boldsymbol{U}$ is the background flow. We integrate these equations with $N=256^2$ wavepackets. These are initially distributed uniformly across the $2\pi\times2\pi$ spatial domain, and their wavenumbers are distributed randomly along a circle of radius $K_0$. We select $K_0$ such that the intrinsic frequency \eqref{eq:shallow-water-dispersion} satisfies $\omega_0=2f$. 

In order to investigate the long-time dynamics of the waves, we advect the rays using a flow generated by steady-state simulations of the two-layer quasigestrophic equations
\begin{subequations}\label{eq:swqg}
    \begin{gather}
    \p_t q_i + J(\psi_i,q_i) + (-1)^{i-1}U_s \left[\p_x q_i 
    + K_D^2\p_x \psi_i\right] = -\delta_{i2}\mu\nabla^2\psi_i-\nu\nabla^8q_i\\
    q_i = \nabla^2\psi_i + (-1)^{i-1}\frac{1}{2}K_D^2(\psi_1-\psi_2),
    \end{gather}
\end{subequations}
forced by a baroclinically-unstable background shear $U_s$ and dissipated by linear drag with coefficient $\mu$. Here $i = 1,2$ denotes the upper or lower layer, respectively, $\psi_i$ is the streamfunction of the $i$-th layer, $J(f, g)=f_xg_y-f_yg_x$ is the Jacobian operator, and $\delta_{ij}$ is the Kronecker delta. The model is simulated using the same resolution, same methods for setting the time step and hyperviscous coefficient, and same pseudo-spectral Julia package as for the direct simulations of the rotating shallow water simulations.  

We generate a range of amplitudes by varying the parameters of the quasigeostrophic simulation to achieve different steady-state energies. For a given background shear $U_s$ and drag $\mu$, the energies are well predicted by vortex-gas scaling theory \citep{GalletFerrari2020}. Across all simulations, we hold $U_s/\mu$ constant, which results in a fixed eddy mixing length, thus ensuring the same scale-separation between the wave-packets and background flow even as the flow strength changes.  Moreover, with a fixed deformation radius and mixing length, vortex-gas scaling theory predicts that the mean eddy velocity $\langle\boldsymbol{u}_\text{balance}\rangle$ scales linearly with $U_s$. Thus, to achieve a range of flow magnitudes, we vary $U_s$ and $\mu$ together, holding their ratio fixed. We apply the constant-enstrophy dissipation method described in Section~\ref{sec:dns}, so $\nu$ does not vary with the amplitude of the flow.

The  mean flow for the ray tracing simulation is the quasigeostrophic baroclinic eddy velocity $\boldsymbol{U}(\boldsymbol{x},t) = (-\psi_y, \psi_x)$, where $\psi = \frac{1}{2}(\psi_1-\psi_2)$ is the baroclinic streamfunction. We use the baroclinic streamfunction as it is affected by the deformation radius, similar to the balanced flow in the rotating shallow water system. The amplitude of this flow determines the effective Froude number, $\textup{Fr}=\langle\boldsymbol{U}\rangle/c$, of the ray tracing simulations.

Each simulation is spun up until it reaches its statistically steady state, after which wavepackets are evolved --- simultaneously with the quasigeostrophic flow --- using the ray tracing equations \eqref{eq:raytracing}. The mean velocity and its gradients are linearly interpolated to evaluate them at the continuous positions of the wavepackets. Each wavepacket begins with some fixed constant action that stays constant as it evolves by the ray tracing equations \eqref{eq:raytracing}, which permits an energy spectrum estimate from a collection of wavepackets using $a = E/\omega$. As action is constant for a given packet, the action spectrum as a function of $\omega$ is simply the distribution of wavepackets in frequency space. We can convert from wavenumber to frequency using the shallow water dispersion relation \eqref{eq:shallow-water-dispersion}. 

\begin{figure}
    \centering
    \begin{subfigure}{0.45\linewidth}
        \centering
        \includegraphics[width=\textwidth]{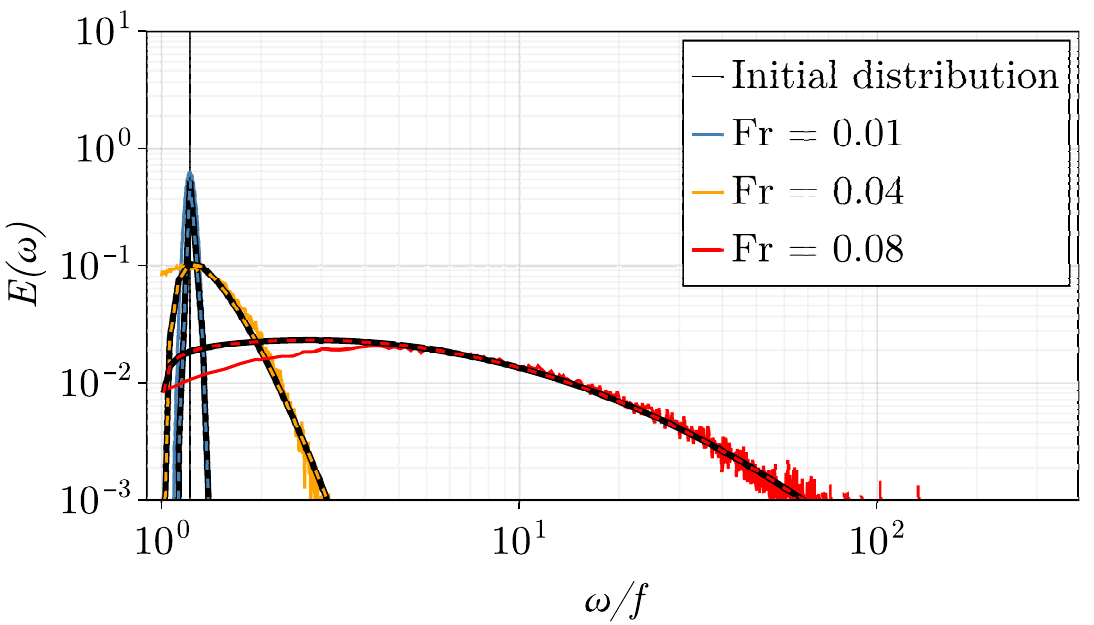}
        \caption{}
        \label{fig:frequency-spreading-a}
    \end{subfigure}
    \begin{subfigure}{0.45\linewidth}
        \centering
        \includegraphics[width=\textwidth]{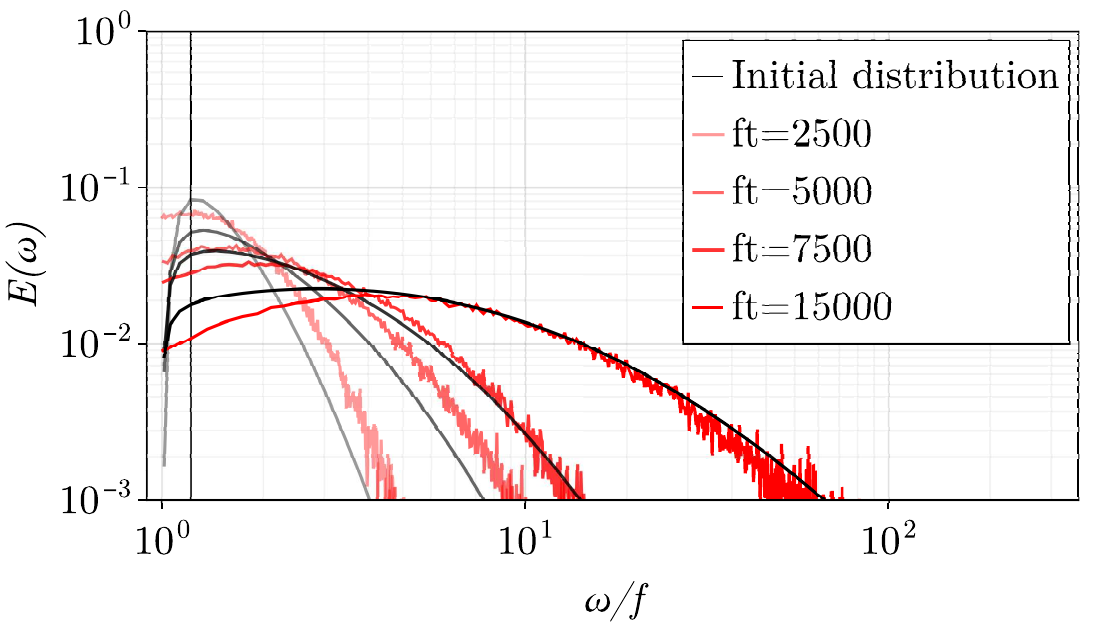}
        \caption{}
        \label{fig:frequency-spreading-in-time}
    \end{subfigure}
    \caption{(\subref{fig:frequency-spreading-a}) Spectra from ray tracing simulation for three background flows with varying amplitudes, taken at the same moment in time ($ft=15000$). The coloured lines show three different simulations with different Froude numbers. The dashed lines show the theoretical energy spectrum computed from a delta function initial condition \eqref{eq:action-diffusion-solution} by fitting the $b$ parameter to the simulation data. (\subref{fig:frequency-spreading-in-time}) Energy spectrum for a single ray tracing simulation as a function of time. The red lines of increasing opacity show the time evolution of the energy spectrum, while the solid lines display the theoretical spectrum using the estimate for $b$ from the left figure.}
    \label{fig:frequency-spreading}
\end{figure}

We run ten ray tracing simulations where we vary the Froude number from $0.01$ to $0.1$, run to the same final time, $ft = 8000$. Figure \ref{fig:frequency-spreading} shows the estimated energy spectra for three of these simulations. As the strength of the mean flow increases, the energy spectrum spreads out further, and its slope becomes shallower. For very small Froude numbers, the packets barely spread from their initial distribution. 
%

\section{Numerical calculations of radial diffusion}\label{sec:analysis}

The energy spectra from the ray tracing simulations suggest a strong dependence of the degree of frequency spreading on the Froude number of the background flow. In this section, define a measure for the strength of frequency spreading using induced-diffusion limit of the wave action equation. We compare this measure of radial diffusivity strength for several flows with varying background flow Froude number, using both a realistic background flow and the synthetic flow used in DBS to explicitly characterise the dependence of radial diffusivity strength on Froude number.

For small-amplitude mean flow ($Fr\ll 1$), the slow evolution of the wave action equation is approximated by the advection-diffusion equation
\begin{equation}\label{eq:action-evolution-diffusion}
    \partial_t a + \bnabla_{\boldsymbol{k}}\omega(\boldsymbol{k})\bcdot\bnabla_{\boldsymbol{x}}a =\bnabla_{\boldsymbol{k}}\bcdot(\mathsfbi{D}\bcdot\bnabla_{\boldsymbol{k}}a),
\end{equation}
where $\mathsfbi{D}$ is a symmetric diffusivity tensor that can be computed asymptotically \citep{kafiabad_diffusion_2019,Dong_Bühler_Smith_2020}. In polar coordinates for wavenumber space such that d$\boldsymbol{k}=K\mathrm{d}K\,\mathrm{d}\theta$, the radial action spectrum is \begin{equation}
    \mathcal{A}(K, t) = \int_0^{2\pi}\int_{\mathbb{R}^2}Ka(\mathbf{x},K,\theta,t)\,\mathrm{d}\mathbf{x}\,\mathrm{d}\theta.
\end{equation}
Hence $\int_0^\infty \mathcal{A}(K,t)\,\mathrm{d}K$ is the total action of all wavepackets, which is time-independent.

We assume a nondivergent mean flow with stationary, isotropic turbulent statistics. 
Integrating \eqref{eq:action-evolution-diffusion} over space and angle results then gives
\begin{equation}\label{eq:radial-action-solvable}
    \partial_t\mathcal{A} = \partial_K\left(KD_{KK}(K)\partial_K\left(\frac{\mathcal{A}}{K}\right)\right),
\end{equation}
%
%
where
\begin{equation}\label{eq:radial-diffusivity}
    D_{KK}(K) = \frac{K^2}{2(2\pi)^2}\int_0^{2\pi}\int_0^\infty q^5\hat{C}(q, -qc_g(K)\cos\eta)\sin^2\eta\cos^2\eta\,\textup{d}q\,\textup{d}\eta.
\end{equation}
Here $c_g(K)$ is the group velocity magnitude at wavenumber $K$, and $\hat{C}(q,\sigma)\geq0$ is the Fourier transform in space and time of the streamfunction autocorrelation
\begin{equation}
    C(r,\tau) = \mathbb{E}\left[\psi(x,y,t)\psi(x+r_x,y+r_y,t+\tau)\right], \quad r=\sqrt{r_x^2+r_y^2}.
\end{equation}
This is well-defined as $\psi$ is stationary, isotropic, and homogeneous.

The shallow water dispersion relationship \eqref{eq:shallow-water-dispersion} is approximately non-dispersive for wavenumbers larger than $K_D$, and hence high frequencies. Since we are interested in spreading to higher frequencies, we make the non-dispersive approximation such that $\omega(\boldsymbol{k})\approx c K$ and $c_g(K)\approx c$. In this case, the entire influence of the mean flow is now encapsulated by a single constant number, 
\begin{equation}
    b=D_{KK}/K^2,
\end{equation} which has the dimension of an inverse time scale. In other words, \textit{all} possible mean flows satisfying the above assumptions lead to identical behaviour in \eqref{eq:radial-action-solvable} if the dynamics is measured in non-dimensional time $bt$.

\subsection{Analytic solution and diffusion time scale}

With initial condition $\mathcal{A}(K, 0) = \mathcal{A}_0K_0\delta(K - K_0)$  and boundary conditions $\mathcal{A}(0,t)=\mathcal{A}(\infty,t)=0$, the change of variables $\mathcal{A} = e^{-bt}\mathcal{B}$ and $K=e^{\sqrt{b}z}$ turns \eqref{eq:radial-action-solvable} into a solvable one-dimensional heat equation. This yields
\begin{equation}\label{eq:action-diffusion-solution}
    \mathcal{A}(K,t) = \frac{\mathcal{A}_0}{\sqrt{4\pi bt}}\exp\left(-bt-\frac{\ln^2\left(\frac{K}{K_0}\right)}{4bt}\right).
\end{equation}
In the rotating shallow water ray tracing setting, it is trivial to convert between wavenumber and frequency and between action and energy because a  wave with wavenumber magnitude $K$ has a unique frequency of $\omega(K)$, so $E=a/\omega$ permits an energy spectrum computed from the action spectrum. The energy spectrum is computed by weighting the action by the corresponding dispersion frequency. Figure \ref{fig:frequency-spreading} shows solutions \eqref{eq:action-diffusion-solution} in energy as a function of frequency. At fixed $K > K_0$, this spectrum peaks in time when
\begin{equation}\label{eq:time-to-max}
    bt_\star = -\frac{1}{4}+\frac{1}{4}\sqrt{1+16\ln^2\left(\frac{K}{K_0}\right)}
\end{equation}
This grows slowly with $K/K_0$, e.g., diffusion across a decade $K/K_0=10$ is already achieved at $bt_\star\approx 2$. This justifies viewing $1/b$ as the effective diffusion time scale in a quantitative fashion. From the solution to the action equation \eqref{eq:action-diffusion-solution} and the plots in Figure \ref{fig:frequency-spreading-a}, it is clear that as the Froude number increases, frequency spreads faster. In the next section, we quantify this spreading by diagnosing $D_{KK}=bK^2$ as a function of the Froude number. 

\subsection{Numerical estimation of the diffusivity }

\begin{figure}
    \centering
    \begin{subfigure}[t]{0.45\linewidth}
        \centering
        \includegraphics[width=0.8\textwidth]{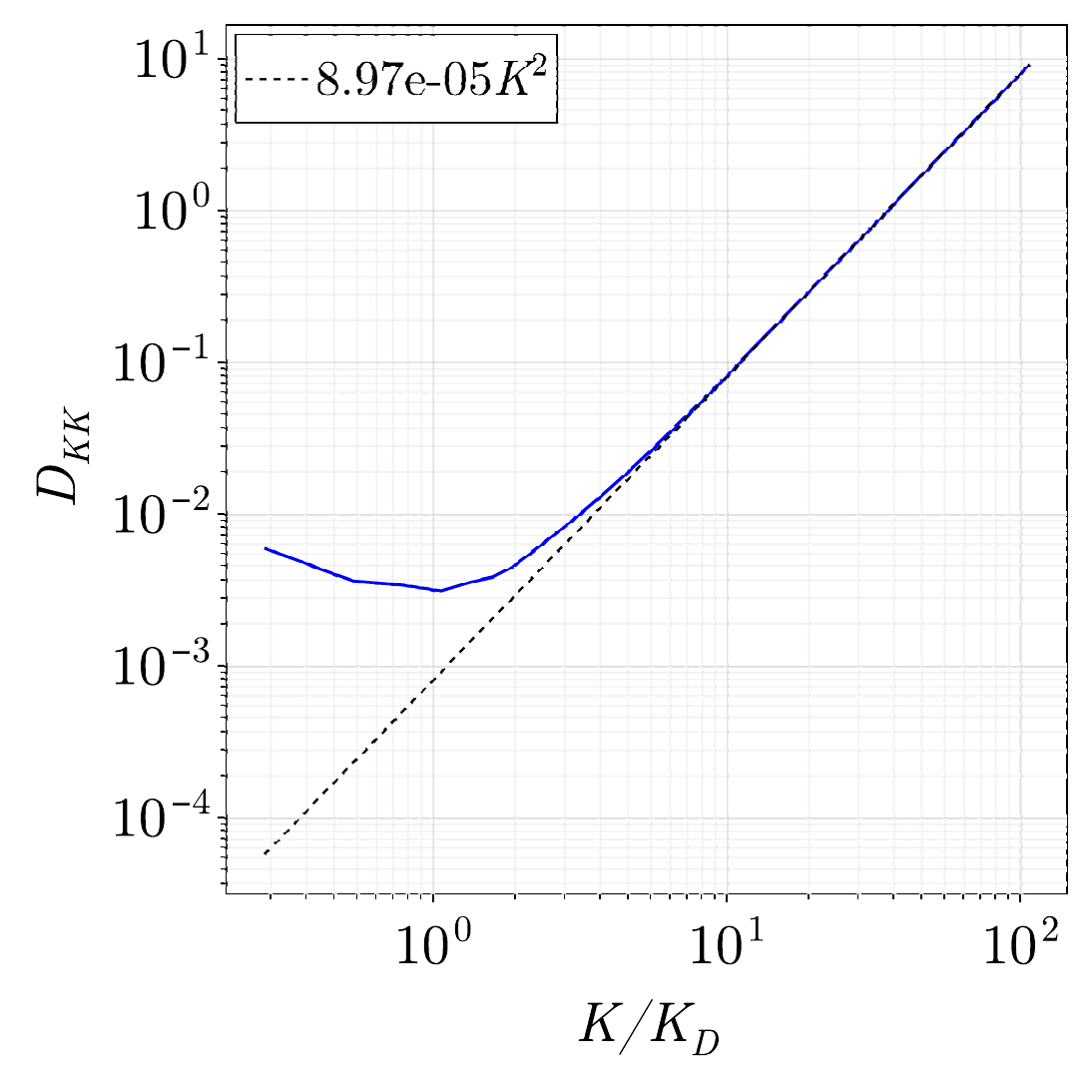}
        \caption{}
        \label{fig:radial-diffusivity-estimate}
    \end{subfigure}
    \begin{subfigure}[t]{0.45\linewidth}
        \centering
        \includegraphics[width=0.8\textwidth]{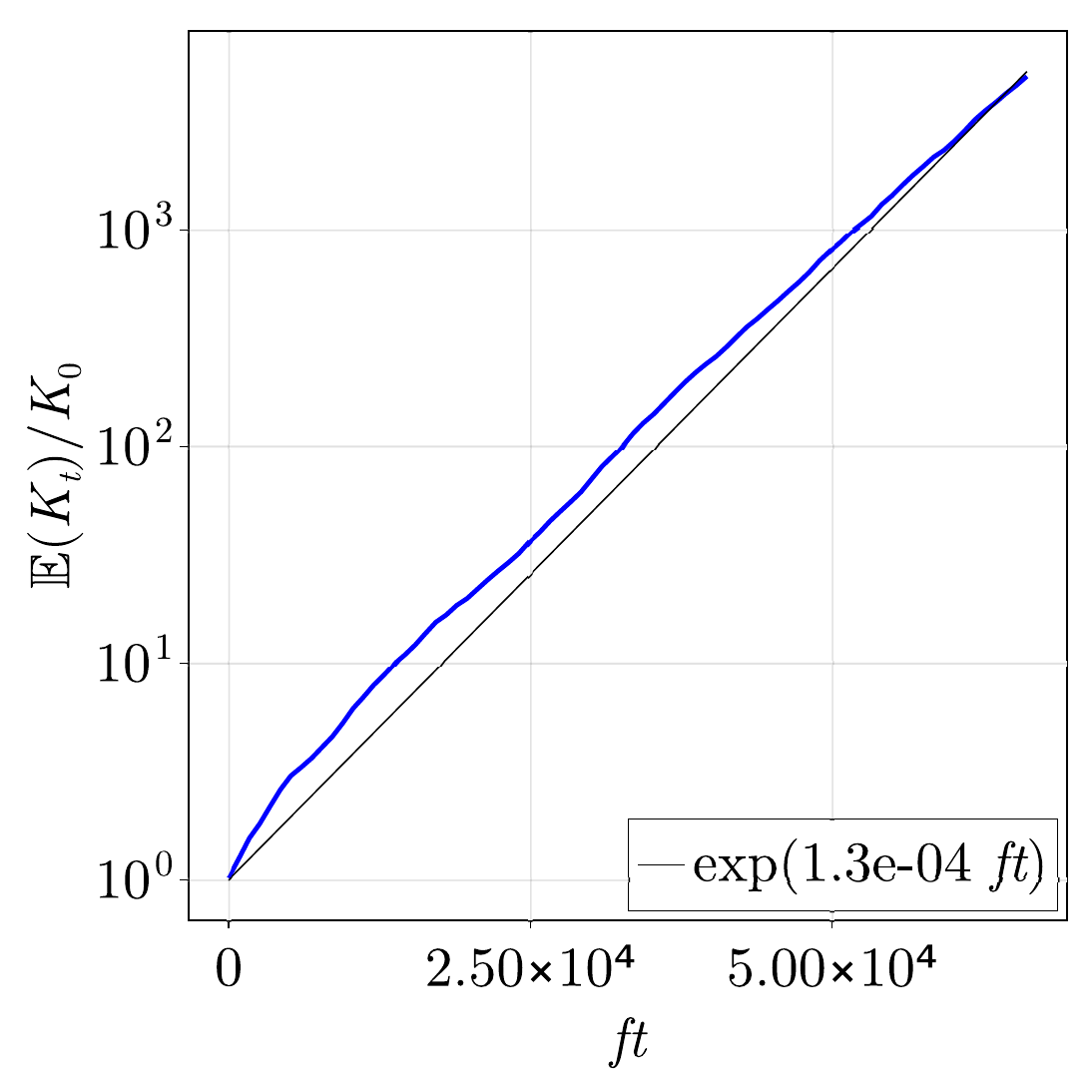}
        \caption{}
        \label{fig:log_mean_gbm}
    \end{subfigure}
    \begin{subfigure}[t]{0.5\linewidth}
        \centering
        \includegraphics[width=0.8\linewidth]{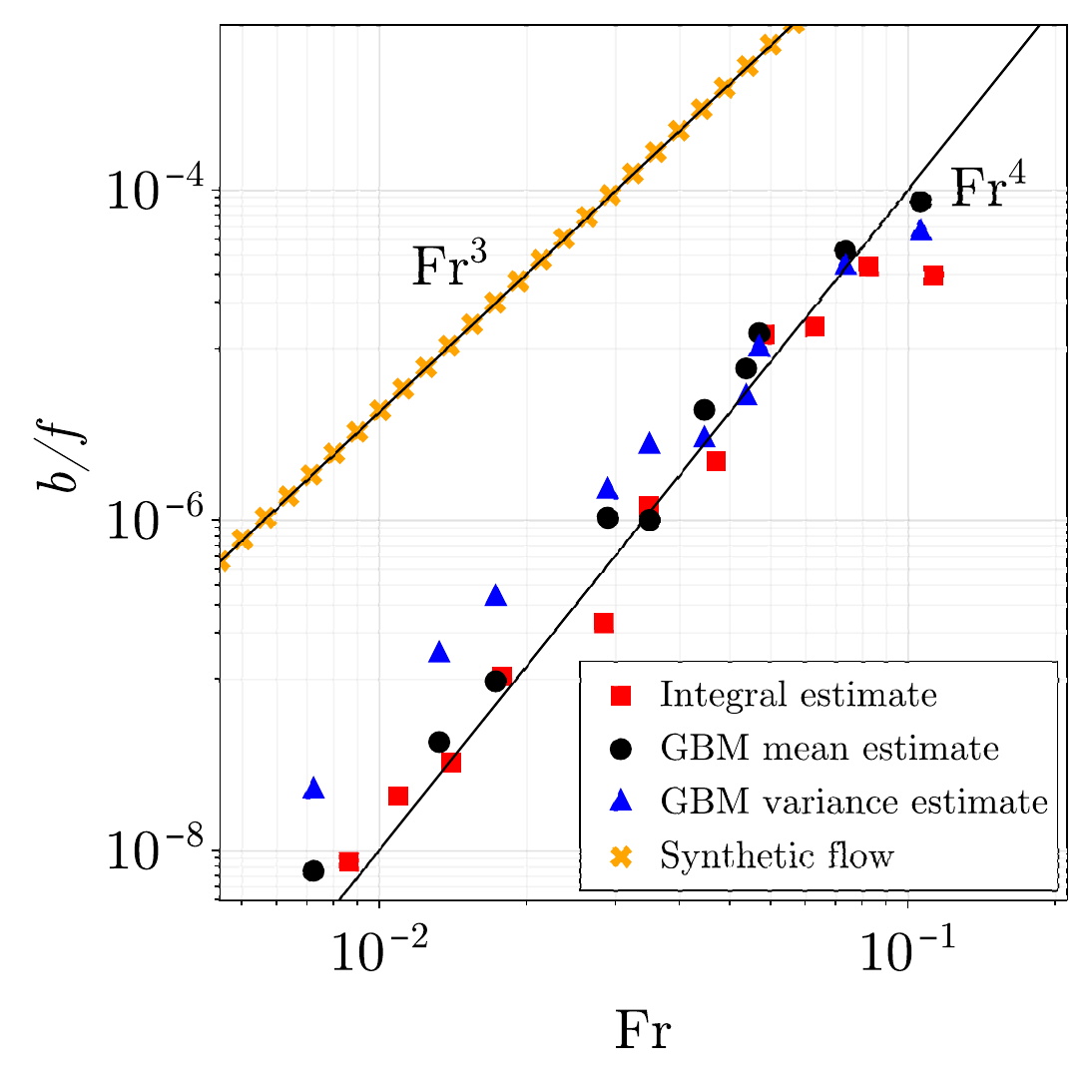}
        \caption{}
        \label{fig:alpha-b-dependence}
    \end{subfigure}
    \caption{(\subref{fig:radial-diffusivity-estimate}) Estimated radial diffusivity by using the flow from a two-layer QG simulation with $Fr = 0.04$ to compute $D_{KK}$ from direct numerical integration of \eqref{eq:radial-diffusivity}. (\subref{fig:log_mean_gbm}) Time-evolution of the mean horizontal wavenumber of wavepackets in a ray tracing simulation using background flow with $Fr = 0.10$. The dashed line is the theoretical growth of a geometric Brownian motion process computed from \eqref{eq:log-mean-gbm}. (\subref{fig:alpha-b-dependence}) The radial diffusivity strength $b$ as a function of the Froude number of the background quasigeostrophic flow, computed in three ways: one from directly computing the diffusivity integral \eqref{eq:radial-diffusivity}, and two from estimating the mean and variance by treating the ray tracing solutions as a geometric Brownian motion process, computed from both parts of \eqref{eq:log-mean-gbm}, respectively. The radial diffusivity using the synthetic flow from DBS is plotted in orange crosses. }
\end{figure}

The radial diffusivity coefficient $D_{KK}=bK^2$ can be estimated from the time-evolving baroclinic quasigeostrophic velocity generated by simulating \eqref{eq:swqg} at a range of flow strengths (used to vary the Froude number). Using high spatial- and temporal-resolution output from a simulation to compute $\hat{C}(q,\sigma)$, and the full shallow water dispersion relationship, we numerically integrate the radial diffusivity integral \eqref{eq:radial-diffusivity}.
The radial diffusivity computed this way for a simulation with a small Froude number ($Fr=0.04$) is shown in Figure \ref{fig:radial-diffusivity-estimate}. As expected, for scales where shallow water waves are approximately non-dispersive, the diffusivity scales as $K^2$, rationalising an investigation of how the prefactor $b$ depends on the Froude number.

\begin{figure}[t!]
    \centering
    \begin{subfigure}[t]{0.24\textwidth}
        \centering
        \includegraphics[width=\linewidth]{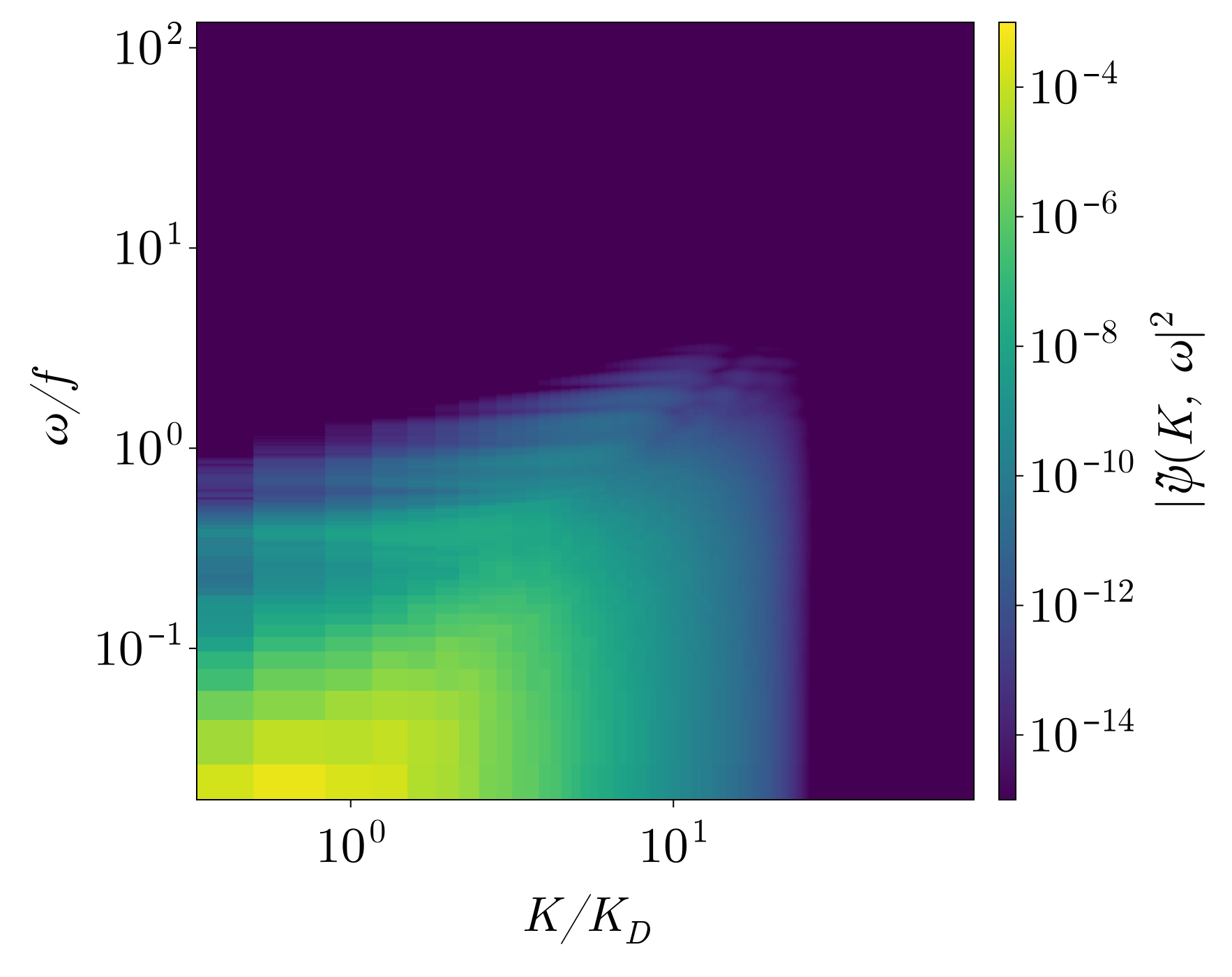}
        \caption{}
        \label{fig:swqg-autocorrelation}
    \end{subfigure}
    \begin{subfigure}[t]{0.24\textwidth}
        \centering
        \includegraphics[width=\linewidth]{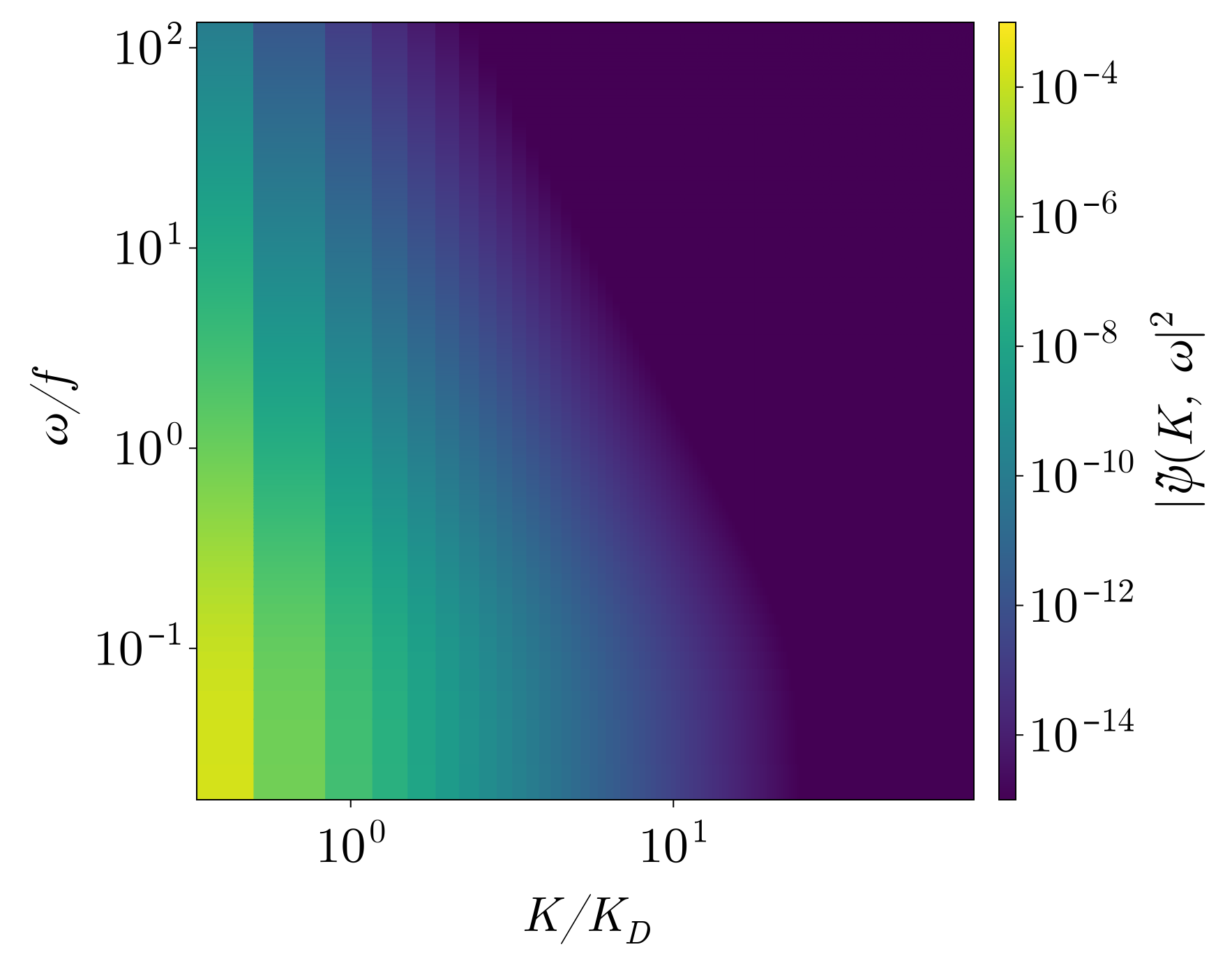}
        \caption{}
        \label{fig:synthetic-autocorrelation}
    \end{subfigure}
    \begin{subfigure}[t]{0.24\textwidth}
        \centering
        \includegraphics[width=0.9\textwidth]{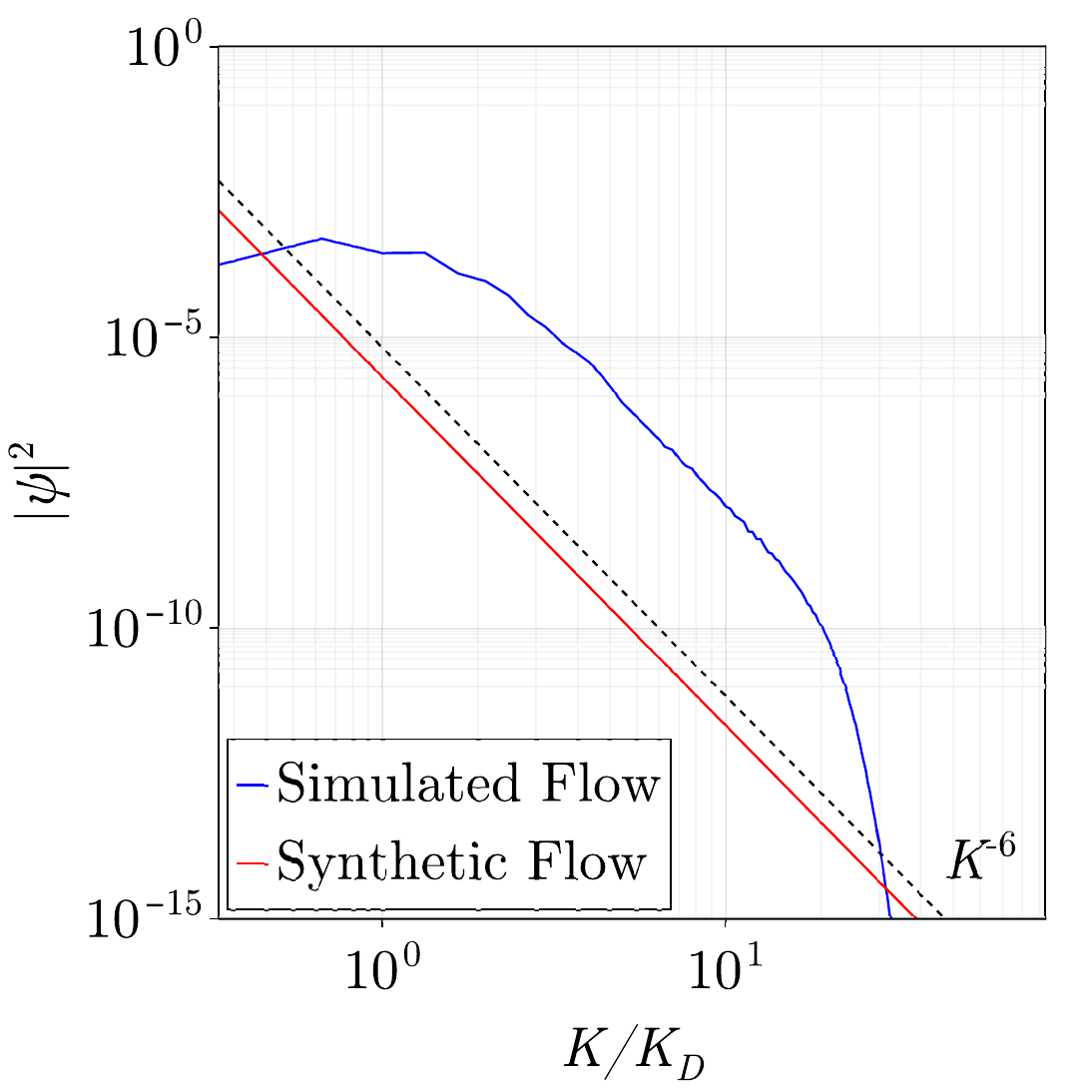}
        \caption{}
        \label{fig:wavenumber-comparison}
    \end{subfigure}
    \begin{subfigure}[t]{0.24\textwidth}
        \centering
        \includegraphics[width=0.9\textwidth]{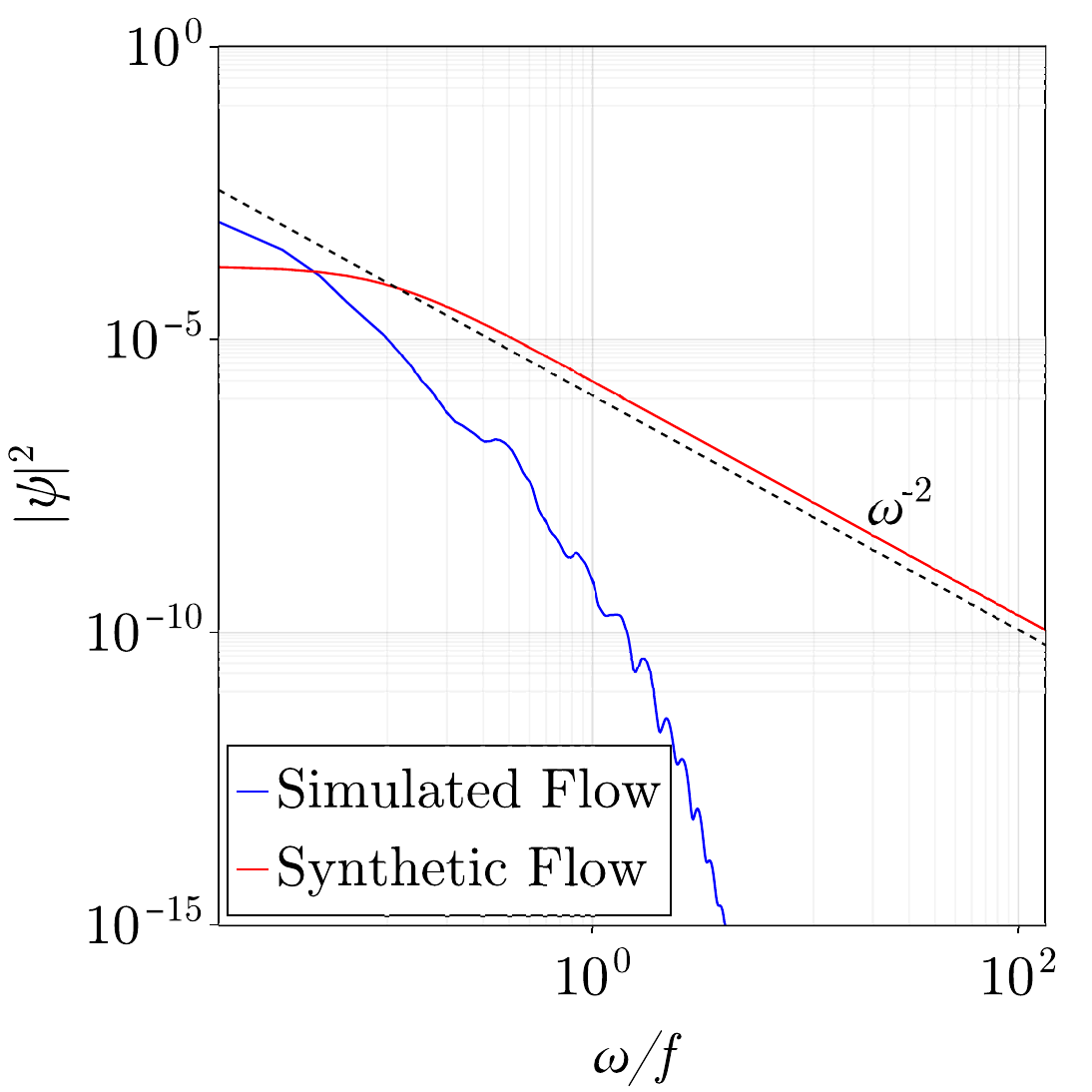}
        \caption{}
        \label{fig:frequency-comparison}
    \end{subfigure}
    \caption{(\subref{fig:swqg-autocorrelation}), (\subref{fig:synthetic-autocorrelation}) Wavenumber-frequency spectra for the simulated quasigeostrophic flow and synthetic flow used in DBS, respectively. The synthetic spectra are defined by an Ornstein-Uhlenbeck process and a fixed power law wavenumber spectrum \eqref{eq:dong-synthetic-spectrum} with spectral slope $-6$. (\subref{fig:wavenumber-comparison}), (\subref{fig:frequency-comparison}) Marginal distributions in wavenumber and frequency, respectively. The marginal wavenumber spectra show the same $k^{-6}$ decay for both synthetic and simulated flows. On the other hand, the simulated flow has a much steeper drop-off in marginal frequency than the synthetic flow.}
    \label{fig:autocorrelation-comparison}
\end{figure}

In order to better understand the diffusivity prefactor, $b$, we compare the statistics of the quasigeostrophic mean flow to those of the synthetic flow used by DBS.    Figure~\ref{fig:swqg-autocorrelation} shows the wavenumber-frequency spectrum of $\vert \hat{\psi}(K,\omega)\vert^2$ for the $Fr = 0.04$ quasigeostrophic flow. By the Wiener--Khinchin theorem, the spectrum can also be computed from the autocorrelation, $\hat{C}(q,\sigma)$.  Figure~\ref{fig:synthetic-autocorrelation} shows the power spectrum for the stochastic synthetic flow used in DBS,
\begin{equation}\label{eq:dong-synthetic-spectrum}
    \hat{C}_{\mathrm{synthetic}}(q, \sigma) = BU_0^2q^{-6}\frac{2\alpha}{\sigma^2+\alpha^2},
\end{equation}
where $B$ is a normalisation constant, $U_0$ is the amplitude of the mean flow, and $\alpha$ is the time dependence of the OU process, set to match the mean eddy turnover time. The exponent $-6$ ensures that the spectral slope of the energy is $k^{-3}$.

While their wavenumber spectra appear similar, the frequency spectrum of the realistically simulated flow is significantly steeper than that of the synthetic flow. We shall see below that this difference leads to different scalings of $b$ with $Fr$.



\subsection{Estimation of the diffusivity by analogy with the Fokker--Planck equation}

Radial diffusivities can also be estimated from the simulations of the ray tracing equations \eqref{eq:raytracing}. Notice that the radial action equation \eqref{eq:radial-action-solvable} can be written as a Fokker--Planck equation via
\begin{equation}
    {\cal A}_t +(3bK{\cal A})_K = (bK^2{\cal A})_{KK}.
\end{equation}
Hence ${\cal A}(K,t)\geq0$ can be viewed as the probability density function for one-dimensional geometric Brownian motion in radial wavenumber $K$ \citep[e.g.,][]{gardiner1985handbook}. Specifically, if $K_t$ is the random wavenumber at time $t$ then its evolution is described by the It\^o stochastic differential equation
\begin{equation}
    \mathrm{d}K_t = 3bK_t\,\mathrm{d}t + \sqrt{2b}K_t\,\mathrm{d}W_t,
\end{equation}
which has an explicit solution. In particular, if the random walk starts at $K_0$ then
\begin{equation}\label{eq:log-mean-gbm}
    \mathbb{E}(K_t/K_0) = e^{3bt}\quad\mbox{and}\quad
    \mathrm{Var}(K_t/K_0)=\left(e^{8bt}-e^{6bt}\right).
\end{equation}
Therefore, the radial diffusivity strength $b$ can be estimated from either the mean or the variance of the time-evolving radial wavenumber of the wavepackets. Figure \ref{fig:log_mean_gbm} shows the time dependence of the log-mean radial wavenumber from a ray tracing simulation, along with the scaling predicted from geometric Brownian motion theory \eqref{eq:log-mean-gbm}. Figure \ref{fig:alpha-b-dependence} shows the dependence of the radial diffusivity on the Froude number of the mean flow, based on several numerical simulations of the two-layer QG equations \eqref{eq:swqg}, and simulations of the ray tracing equations \eqref{eq:raytracing} using the method above. For small Froude numbers, there is a clear $b\sim\textrm{Fr}^4$ relationship. This implies that as the amplitude of the mean flow decreases, the rate of diffusion drops off more quickly than for the synthetic flow, which follows $b\sim\textrm{Fr}^3$. A key take-away from Figure \ref{fig:radial-diffusivity-estimate} is that, even for reasonably large Froude numbers, the realistic flow results in an order of magnitude decrease in radial spreading.

\section{Discussion}

We have defined a measure of the degree of wave energy spreading based on the induced-diffusion limit of the wave action equation in a two-dimensional setting. Through numerical simulations of the ray tracing equations, we demonstrate that the degree of frequency spreading is an order of magnitude smaller with a realistic model of a turbulent mean flow than compared to the synthetic mean flow considered by DBS. This is the main result of this paper, which tightens the gap between the two-dimensional setting, where constant frequency surfaces are compact, so frequency spreading is more effective, and the three-dimensional setting considered in \cite{Cox_Kafiabad_Vanneste_2023}, where constant frequency surfaces are not compact and severely limit frequency spreading.

Our results suggest that both wave triad and wave-mean flow interactions are necessary to achieve broadband frequency spreading. For example, as shown in Figure \ref{fig:rsw-msw-comparison}, without the presence of shocks, the spread of wave energy in frequency space is muted. This finding aligns with other work exploring wave-mean flow interactions in the induced diffusion regime. \cite{kafiabad_diffusion_2019, dong_geostrophic_2023, Cox_Kafiabad_Vanneste_2023} explore the three-dimensional setting and find that even with time dependence, wave action does not spread too much off of cones of constant frequency unless the induced diffusion assumptions break down, as in \cite{dong_geostrophic_2023}. \cite{Cox_Kafiabad_Vanneste_2025} explores other mechanisms of wavenumber diffusion besides Doppler shifting, such as variations in the depth of a shallow water layer by bottom topography. However, in the regime we consider, the strength of these effects is comparable to that of Doppler shifting. 

We find that for a realistic two-dimensional mean flow, the strength of radial wavenumber spreading, and hence frequency spreading, scales like $b\sim Fr^4$, while for synthetic flows, this dependence is $b\sim Fr^3$. This is closer to the lack of spreading seen in the three-dimensional setting; for small-Froude-number flows, it takes much longer for wave energy to spread with realistic turbulence than a synthetic flow. As an application to ocean flows, consider wave energy injected near the Coriolis frequency corresponding to some action $\mathcal{A}_0$, and dissipated near the Brunt-Vaisala frequency. In the ocean, the ratio of these frequencies is around $O(100)$, as is the ratio of the corresponding wavenumbers corresponding to these frequencies by the shallow water dispersion relationship \eqref{eq:rsw}. According to the analytical solution of the radial action equation \eqref{eq:action-diffusion-solution}, for a wavenumber ratio $K/K_0 = 100$, the solution reaches a maximum value of $\mathcal{A}_\star = 1.9\cdot10^{-3}\mathcal{A}_0$ at time $bt_\star\approx 4$. For a mean flow with Froude number on the order $O(0.1)$, corresponding with mesoscale eddies, our scaling gives a value of $b$ on the order of $O(10^{-4} f)$. This gives a $t_\star$ on the order of decades; much longer than the lifetime of internal waves. 

Induced diffusion theory only models one-way wave-mean flow interactions; waves feel the effect of a mean flow, but the mean flow does not feel the effect of the waves. We suggest that induced-diffusion wave mean-flow interactions alone are unlikely to be the primary driver of downscale energy flux. Additional processes such as wave capture or wave–wave interactions are therefore likely required to explain the broadband internal-wave spectrum observed in the atmosphere and ocean.

\begin{bmhead}[Acknowledgements.]The computational requirements for this work were supported by the NYU IT High Performance Computing resources and personnel.
\end{bmhead}
\begin{bmhead}[Funding.]ND was supported by an NSF Graduate
Research Fellowship under Grant No. DGE-2039655. OB acknowledges financial support under NSF grant DMS-2406767. 
\end{bmhead}

\begin{bmhead}[Declaration of interests.]The authors report no conflict of interest.
\end{bmhead}
 \begin{bmhead}[Data availability statement.]The data and code that support the findings of this study are openly available at \url{https://github.com/ndefilippis/JuliaRaytracingSW}
\end{bmhead}
\begin{bmhead}[Author ORCIDs.]N. DeFilippis, https://orcid.org/0009-0001-6574-5404; O. B\"uhler, https://orcid.org/0000-0002-4914-8546; K.S. Smith, https://orcid.org/0000-0003-0740-3067 
\end{bmhead}
%
\appendix
\begin{appen}


\end{appen}\clearpage

\bibliographystyle{jfm}
\bibliography{references}

@article{Dong_Bühler_Smith_2020, title={Frequency diffusion of waves by unsteady flows}, volume={905}, DOI={10.1017/jfm.2020.837}, journal={Journal of Fluid Mechanics}, author={Dong, Wenjing and Bühler, Oliver and Smith, K. Shafer}, year={2020}, pages={R3}}

@article{dong_geostrophic_2023,
	title = {Geostrophic {Eddies} {Spread} {Near}-{Inertial} {Wave} {Energy} to {High} {Frequencies}},
	url = {https://journals.ametsoc.org/view/journals/phoc/53/5/JPO-D-22-0153.1.xml},
	doi = {10.1175/JPO-D-22-0153.1},
    journal   = {Journal of Physical Oceanography},
    publisher = {American Meteorological Society},
	language = {en},
	urldate = {2024-07-23},
	author = {Dong, Wenjing and Bühler, Oliver and Smith, K. Shafer},
	month = may,
	year = {2023},
	note = {Section: Journal of Physical Oceanography},
}

@article{garrett_space-time_1972,
	title = {Space-{Time} scales of internal waves},
	volume = {3},
	issn = {0016-7991},
	url = {https://doi.org/10.1080/03091927208236082},
	doi = {10.1080/03091927208236082},
	number = {3},
	urldate = {2024-07-23},
	journal = {Geophysical Fluid Dynamics},
	author = {Garrett, Christopher and Munk, Walter},
	month = may,
	year = {1972},
	note = {Publisher: Taylor \& Francis
\_eprint: https://doi.org/10.1080/03091927208236082},
	pages = {225--264},
}

@article{kafiabad_diffusion_2019,
	title = {Diffusion of inertia-gravity waves by geostrophic turbulence},
	volume = {869},
	issn = {0022-1120, 1469-7645},
	url = {https://www.cambridge.org/core/journals/journal-of-fluid-mechanics/article/abs/diffusion-of-inertiagravity-waves-by-geostrophic-turbulence/31E102FB36D4066476A6A5862B4BFCD8},
	doi = {10.1017/jfm.2019.300},
	urldate = {2025-01-15},
	journal = {Journal of Fluid Mechanics},
	author = {Kafiabad, Hossein A. and Savva, Miles A. C. and Vanneste, Jacques},
	month = jun,
	year = {2019},
	pages = {R7},
}

@article{thomas_turbulent_2024,
	title = {The turbulent cascade of inertia-gravity waves in rotating shallow water},
	volume = {1000},
	issn = {0022-1120, 1469-7645},
	url = {https://www.cambridge.org/core/journals/journal-of-fluid-mechanics/article/turbulent-cascade-of-inertiagravity-waves-in-rotating-shallow-water/48079097EA9B24151734532C79D98695},
	doi = {10.1017/jfm.2024.854},
	language = {en},
	urldate = {2024-12-03},
	journal = {Journal of Fluid Mechanics},
	author = {Thomas, Jim and Rajpoot, Rajendra S. and Gupta, Prateek},
	month = dec,
	year = {2024},
	pages = {A30},
}

@book{SalmonRick1998Logf,
address = {New York},
edition = {1},
isbn = {0195108086},
language = {eng},
publisher = {Oxford University Press},
title = {Lectures on geophysical fluid dynamics},
year = {1998},
author = {Salmon, Rick},
}

@misc{navid_c_constantinou_2025_17281674,
  author       = {Navid C. Constantinou and
                  Gregory L. Wagner and
                  André Palóczy and
                  Ka Wai HO and
                  Josef Bisits and
                  Julia TagBot and
                  Morten Piibeleht and
                  Tim Besard and
                  Connor Robertson and
                  Vladimir Parfenyev},
  title        = {FourierFlows/FourierFlows.jl: v0.10.7},
  month        = oct,
  year         = 2025,
  publisher    = {Zenodo},
  version      = {v0.10.7},
  doi          = {10.5281/zenodo.17281674},
  url          = {https://doi.org/10.5281/zenodo.17281674},
  swhid        = {swh:1:dir:487609a2e7c5d224b761757845b47a8b9172f754
                   ;origin=https://doi.org/10.5281/zenodo.1161724;vis
                   it=swh:1:snp:2ed46cd70b49843623be0cc9fe5d32055ce52
                   c7c;anchor=swh:1:rel:d7b8d66514f2c56d1484224827193
                   316b58593ee;path=FourierFlows-
                   FourierFlows.jl-0cd5f7a
                  },
}

@article { AShallowWaterModelthatPreventsNonlinearSteepeningofGravityWaves,
      author = "Oliver Bühler",
      title = "A Shallow-Water Model that Prevents Nonlinear Steepening of Gravity Waves",
      journal = "Journal of the Atmospheric Sciences",
      year = "1998",
      publisher = "American Meteorological Society",
      address = "Boston MA, USA",
      volume = "55",
      number = "17",
      doi = "10.1175/1520-0469(1998)055<2884:ASWMTP>2.0.CO;2",
      pages=      "2884 - 2891",
      url = "https://journals.ametsoc.org/view/journals/atsc/55/17/1520-0469_1998_055_2884_aswmtp_2.0.co_2.xml"
}

@article{Cox_Kafiabad_Vanneste_2023, title={Inertia-gravity-wave diffusion by geostrophic turbulence: the impact of flow time dependence}, volume={958}, DOI={10.1017/jfm.2023.83}, journal={Journal of Fluid Mechanics}, author={Cox, Michael R. and Kafiabad, Hossein A. and Vanneste, Jacques}, year={2023}, pages={A21}}

@article{ctx40894637160007876,
author = {McComas, C Henry and Bretherton, Francis P},
address = {Richmond, Va. :},
issn = {0148-0227},
journal = {Journal of geophysical research.},
lccn = {80643369},
number = {9},
publisher = {William Byrd Press for John Hopkins Press,},
title = {Resonant interaction of oceanic internal waves},
volume = {82},
year = {1977},
}

@article{Buhler_McIntyre_2005, title={Wave capture and wave–vortex dual}, volume={534}, DOI={10.1017/S0022112005004374}, journal={Journal of Fluid Mechanics}, author={B\"uhler, Oliver and McIntyre, Michael E.}, year={2005}, pages={67–95}}

@article{augier2019shallow,
  title={Shallow water wave turbulence},
  author={Augier, Pierre and Mohanan, Ashwin Vishnu and Lindborg, Erik},
  journal={Journal of Fluid Mechanics},
  volume={874},
  pages={1169--1196},
  year={2019},
  publisher={Cambridge University Press}
}

@book{gardiner1985handbook,
  title={Handbook of stochastic methods},
  author={Gardiner, Crispin W and others},
  volume={3},
  year={1985},
  publisher={springer Berlin}
}

@article{GalletFerrari2020,
author = {Basile Gallet  and Raffaele Ferrari },
title = {The vortex gas scaling regime of baroclinic turbulence},
journal = {Proceedings of the National Academy of Sciences},
volume = {117},
number = {9},
pages = {4491-4497},
year = {2020},
doi = {10.1073/pnas.1916272117},
URL = {https://www.pnas.org/doi/abs/10.1073/pnas.1916272117},
eprint = {https://www.pnas.org/doi/pdf/10.1073/pnas.1916272117}}

@article{Cox_Kafiabad_Vanneste_2025, title={Inhomogeneity-induced wavenumber diffusion}, volume={1007}, DOI={10.1017/jfm.2025.23}, journal={Journal of Fluid Mechanics}, author={Cox, Michael R. and Kafiabad, Hossein A. and Vanneste, Jacques}, year={2025}, pages={A15}}

@article { TheoreticalInterpretationofAtmosphericWavenumberSpectraofWindandTemperatureObservedbyCommercialAircraftDuringGASP,
      author = "K. S.  Gage and G. D.  Nastrom",
      title = "Theoretical Interpretation of Atmospheric Wavenumber Spectra of Wind and Temperature Observed by Commercial Aircraft During GASP",
      journal = "Journal of Atmospheric Sciences",
      year = "1986",
      publisher = "American Meteorological Society",
      address = "Boston MA, USA",
      volume = "43",
      number = "7",
      doi = "10.1175/1520-0469(1986)043<0729:TIOAWS>2.0.CO;2",
      pages=      "729 - 740",
      url = "https://journals.ametsoc.org/view/journals/atsc/43/7/1520-0469_1986_043_0729_tioaws_2_0_co_2.xml"
}

@article { InterpretingEnergyandTracerSpectraofUpperOceanTurbulenceintheSubmesoscaleRange1200km,
      author = "Jörn Callies and Raffaele Ferrari",
      title = "Interpreting Energy and Tracer Spectra of Upper-Ocean Turbulence in the Submesoscale Range (1–200 km)",
      journal = "Journal of Physical Oceanography",
      year = "2013",
      publisher = "American Meteorological Society",
      address = "Boston MA, USA",
      volume = "43",
      number = "11",
      doi = "10.1175/JPO-D-13-063.1",
      pages=      "2456 - 2474",
      url = "https://journals.ametsoc.org/view/journals/phoc/43/11/jpo-d-13-063.1.xml"
}

\end{document}